# A Family of Robust Generalized Adaptive Filters and Application for Time-series Prediction

Yi Peng, Haiquan Zhao, *Senior Member, IEEE*, Jinhui Hu

*Abstract*—The continuous development of new adaptive filters (AFs) based on novel cost functions (CFs) is driven by the demands of various application scenarios and noise environments. However, these algorithms typically demonstrate optimal performance only in specific conditions. In the event of the noise change, the performance of these AFs often declines, rendering simple parameter adjustments ineffective. Instead, a modification of the CF is necessary. To address this issue, the robust generalized adaptive AF (RGA-AF) with strong adaptability and flexibility is proposed in this paper. The flexibility of the RGA-AF's CF allows for smooth adaptation to varying noise environments through parameter adjustments, ensuring optimal filtering performance in diverse scenarios. Moreover, we introduce several fundamental properties of negative RGA (NRGA) entropy and present the negative asymmetric RGA AF (NARGA-AF) and kernel recursive NRGA AF (KRNRGA-AF). These AFs address asymmetric noise distribution and nonlinear filtering issues, respectively. Simulations of linear system identification and time-series prediction for Chua's circuit under different noise environments demonstrate the superiority of the proposed algorithms in comparison to existing techniques.

*Index Terms*— robust adaptive filter, generalized Gaussian noise, asymmetric noise, kernel method, kernel recursive algorithm

## I. INTRODUCTION

Choosing the appropriate cost function (CF) for different environments is crucial in determining the performance of adaptive filter (AF) algorithms [1]. Algorithms developed through the mean squared error (MSE) criterion, including the least mean square (LMS) algorithm and its improved variants like variable step size LMS (VSSLMS) algorithm [2] and normalized LMS algorithm (NLMS) [3], are extensively utilized in a variety of scenarios due to the smoothness, convexity, and optimality of the MSE criterion under Gaussian assumptions. These applications include system identification, echo cancellation, and channel equalization [4], [5], [6].

However, the AFs using higher-order moments can attain better filtering results compared to algorithms developed via the MSE criterion. Particularly under non-Gaussian conditions, such as light-tailed noise, like uniform or binary noise,

This work was partially supported by National Natural Science Foundation of China (grant: 62171388, 61871461, 61571374).

Haiquan Zhao, Yi Peng and Jinhui Hu are with the School of Electrical Engineering, Southwest Jiaotong University, Chengdu, 610031, China. (e-mail: hqzhao_swjtu@126.com; pengyi1007@163.com ; jhhu_swjtu@126.com).

Corresponding author: Haiquan Zhao.

the least mean fourth (LMF) algorithm [7], which uses the fourth-order error moment as the CF, can achieve faster convergence and lower steady-state MSD. Nevertheless, compared to the LMS algorithm, the stability of the LMF algorithm is diminished. Additionally, the least mean p-power (LMP) algorithm [8], which uses the p-th order absolute moment of the error as its CF, is a more comprehensive approach than the LMF algorithm.

In contrast, algorithms developed using the lower-order error moments, like the sign algorithm (SA) [9], [10], generally do not perform as well in terms of convergence. Nevertheless, when the noise environment contains impulsive noise, the SA algorithm can achieve better performance than the algorithms using higher-order error moments due to its insensitivity to impulsive interference. Many robustness algorithms have been designed to address the performance degradation caused by impulsive noise. These include the least mean logarithmic square (LMLS) [11] and robust least mean logarithmic square (RLMLS) [12] algorithms through the logarithmic CF, the least mean M-estimate (LMM) [13], [14] algorithm using the M-estimation function, the robust mixed-norm (RMN) algorithm [15], [16], [17] based on a mixed CF, and the maximum correntropy criterion (MCC) algorithm [18], [19], [20], [21] based on information theoretic learning.

However, in real-world environments, the noise is not limited to impulsive noise. When the noise adheres to a generalized Gaussian distribution [22], the MCC algorithm, may not perform well due to its reliance on a Gaussian kernel. Therefore, the generalized MCC (GMCC) algorithm [23] was developed, by employing the generalized Gaussian kernel [24], [25]. Several improved robust algorithms [26], [27] have been proposed in recent years to address the issue of high steady-state error in the GMCC algorithm. In cases where the noise distribution is asymmetric [28], which is common in wireless localization [29], symmetric functions like the Gaussian and generalized Gaussian kernels may not adequately approximate the probability density function (PDF) of the error, potentially leading to significant performance degradation. To address this issue, the maximum asymmetric correntropy criterion (MACC) [30] and the generalized MACC (GMACC) [31] algorithms have been proposed.

In addition, conventional AFs cannot solve the problem of chaotic sequence prediction such as time-series prediction. Currently, dynamic neural network (DNN), support vector regression (SVR), and kernel adaptive filtering (KAF) have been applied to time-series prediction [32], [33], [34]. However, both DNN and SVR operate offline, DNN involves a large number of computations during the update process [35], and SVR is suitable for small-sample prediction and tracking. KAF algorithm not only has low computational complexity and simple recursive update form, but also meets the requirements of online prediction. A number of kernel adaptive filtering (KAF) algorithms [36], [37] have been developed for time-series prediction. These algorithms use the kernel method to map the input signal into a high-dimensional feature space, solving the problem of linear indivisibility in low-dimensional space.

Unfortunately, the efficacy of existing CFs and AFs, designed for specific environments, is frequently compromised when the noise environment undergoes a change. Simply adjusting algorithm parameters cannot avoid this performance degradation. Instead, the CF needs to be redesigned. It is therefore necessary to design a CF that is capable of optimal performance in the diverse application scenarios and varying noise environments, which motivates this work. In this paper, a new robust generalized adaptive (RGA) CF with adaptability and flexibility has been proposed. This CF can be regarded as a synthesis of multiple CFs, with the parameters allowing for the selection and combination of the appropriate CFs, thereby leveraging their respective strengths to address diverse environmental challenges. Moreover, when the parameters meet certain conditions, the CF can be viewed as an expansion of the NRGA kernel, possessing properties similar to generalized Gaussian kernel. Similarly, the NRGA kernel can be transformed into other kernels when the parameters are changed. Detailed analysis of the NRGA entropy is provided in this paper. To address asymmetric error distributions, the NARGA algorithm has been developed. For the time-series prediction problem of Chua's circuit, we further develop the KRNRGA algorithm using the kernel method. Finally, the computer simulations demonstrated the excellent performance of the proposed algorithms under various conditions. The main contributions of this paper are summarized as follows:

1) The RGA-AF is proposed in this paper, which can adapt to different noise environments by adjusting parameters. It achieves better performance compared to other competing algorithms under various noise distributions. Additionally, we analyzed the convergence of the RGA algorithm, calculated its steady-state MSD, and finally verified the correctness of the calculations by simulation.

2) The properties of the NRGA entropy have been analyzed in detail, showing that the NRGA entropy has similar properties to the generalized maximum correntropy under certain parameters.

3) To address the issue of performance degradation under asymmetric noise distribution, an asymmetric NARGA kernel is proposed. The NARGA kernel fits the PDF of asymmetric noise distributions very well. Simulations have demonstrated that the NARGA algorithm outperforms other competing algorithms.

4) For the time series prediction problem of Chua's circuit, we propose the KRNRGA algorithm using the kernel method. Simulations have demonstrated that the KRNRGA algorithm outperforms other competing algorithms.

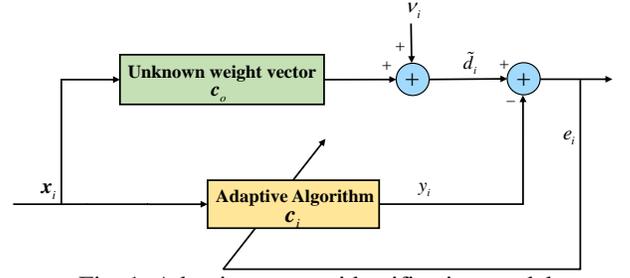

Fig. 1. Adaptive system identification model

A brief description of the subsequent sections in this paper is provided below. Section II briefly reviews the system identification model, and the definition of the general adaptive (GA) function. Section III provides the definitions of the RGA and NRGA functions, as well as some properties of the NRGA entropy and the asymmetric kernel function. Section IV derives the RGA algorithm, the NARGA algorithm and the KRNRGA algorithm in detail. Section V analyzes the convergence and calculates the steady-state MSD of the RGA algorithm. The computational complexity of the different algorithms is given in Section VI. Detailed computer simulations and discussions are provided in the Section VII. The conclusion is presented in Section VIII.

## II. BRIEF REVIEW

### A. System identification model

As depicted in Fig. 1, the output signal for the system identification problem is represented as

$$\tilde{d}_i = c_o^T x_i + v_i \quad (1)$$

where $c_o \in \mathbb{R}^{L \times 1}$ is a vector of unknown weight weights, $i$ defines discrete time, $x_i \in \mathbb{R}^{L \times 1}$ is the input vector, and $v_i$ is the measurement noise.

In Fig.1, the $y_i = c_i^T x_i$ is the actual output signal, and the error $e_i$ can be indicated as

$$e_i = \tilde{d}_i - c_i^T x_i \quad (2)$$

where $c_i \in \mathbb{R}^{L \times 1}$ is the weight vector.

### B. Definition of GA function

In [38], [39], the GA loss function is proposed. By adjusting the parameters of this function, different loss functions such as Cauchy loss, L2 loss, and Welsch loss can be fitted. This function can be represented as

$$P_{GA}(e, \delta, c) = \frac{|\delta - 2|}{\delta} \left( \left( \frac{|e/c|^2}{|\delta - 2|} + 1 \right)^{\frac{\delta}{2}} - 1 \right) \quad (3)$$

where, $\delta \in \mathbb{R}$ is shape parameter, and $c > 0$ is scale parameter.

*Remark 1:* When $\delta$ approaches 0, 2, and $-\infty$, the algorithm derived from the GA function can fit the LMLS, LMS, and MCC algorithm, respectively. Although the GA function can fit some CFs commonly used in the AF domain, it is not applicable to complex environments. This is because, 1) the error order is limited to 2, which prevents the use of higher order error statistics, and 2) the kernel function of the MCC algorithm derived from the GA function is a Gaussian kernel, which is not always the optimal choice.

### III. IMPORTANT DEFINITIONS AND KERNEL METHODS

In this section, several important definitions will be provided to lay the groundwork for the subsequent algorithms proposed. Additionally, detailed analysis of the NRGA entropy is provided in this subsection.

*A. Definition of RGA function*

In an effort to more effectively utilize the higher-order moment information of the error, rather than merely the quadratic error moment, and to extend the Gaussian kernel to the generalized Gaussian kernel, the function in (3) is further generalized to the RGA function, which is defined as

$$P_{RGA}(e,\alpha,\beta,\lambda) = \frac{|\alpha-\beta|}{\alpha}\left(\frac{\lambda|e|^\beta}{|\alpha-\beta|}+1\right)^{\frac{\alpha}{\beta}} - \frac{|\alpha-\beta|}{\alpha} \quad (4)$$

here, $\beta > 0$ and $\alpha \in \mathbb{R}$ are shape parameters, and $\lambda > 0$ is scale parameter. For $\beta = 2$, the RGA function reduces to the GA function.

It is evident that the RGA function is not defined at certain points, which will be discussed as follows

*a)* When $\alpha = \beta$

$$\lim_{\alpha \to \beta} \frac{|\alpha-\beta|}{\alpha}\left(\frac{\lambda|e|^\beta}{|\alpha-\beta|}+1\right)^{\frac{\alpha}{\beta}} - \frac{|\alpha-\beta|}{\alpha} = \frac{\lambda}{\beta}|e|^\beta \quad (5)$$

In this case, (5) can be considered as the LMP algorithm. In particular, for $\beta = 4$, (5) can be seen as the LMF algorithm and for $\beta = 2$, (5) can be seen as the LMS algorithm.

*b)* When $\alpha = 0$

$$\lim_{\alpha \to 0} \frac{|\alpha-\beta|}{\alpha}\left(\frac{\lambda|e|^\beta}{|\alpha-\beta|}+1\right)^{\frac{\alpha}{\beta}} - \frac{|\alpha-\beta|}{\alpha} = \log\left(\frac{\lambda}{\beta}|e|^\beta+1\right) \quad (6)$$

In this case, (6) can be considered as the RLMLS algorithm. Specially, for $\beta = 2$, (6) can be seen as the LMLS algorithm.

*c)* When $\alpha \to -\infty$

$$\lim_{\alpha \to -\infty} \frac{|\alpha-\beta|}{\alpha}\left(\frac{\lambda|e|^\beta}{|\alpha-\beta|}+1\right)^{\frac{\alpha}{\beta}} - \frac{|\alpha-\beta|}{\alpha} = 1-\exp\left(-\frac{\lambda}{\beta}|e|^\beta\right) \quad (7)$$

In this case (7) can be considered as the GMCC algorithm. For $\beta = 2$, (7) can be seen as the MCC algorithm.

*d)* When $\alpha \to \infty$

$$\lim_{\alpha \to +\infty} P_{RGA}(e,\alpha,\beta,\lambda) = \exp\left(\frac{\lambda}{\beta}|e|^\beta\right) - 1 \quad (8)$$

As shown in Fig. 2, when $\alpha \to \infty$, the RGA algorithm fails to converge properly. Consequently, this scenario will not be considered in subsequent analyses.

*Remark 2:* The form of the RGA function can be transformed into the CFs of other existing adaptive algorithms when $\alpha$ and $\beta$ are some special values. In Fig. 3, as $\alpha$ tends to $-\infty$, 0, $\beta$ and $\beta$ takes on special values, we compare the CF and gradient of the RGA with the corresponding algorithms [7], [8], [9], [12], [23], respectively. It is shown that the RGA function, which is derived by generalizing the GA function, is able to efficiently utilize a variety of information about the error, rather than just considering second-order moment of error. Different parameters determine the curves of these CFs, delineating sensitive and insensitive intervals to error variations, thus tuning the robustness of the algorithm. In other words, the robustness of the RGA algorithm can be controlled by tuning the parameters to handle various noise environments.

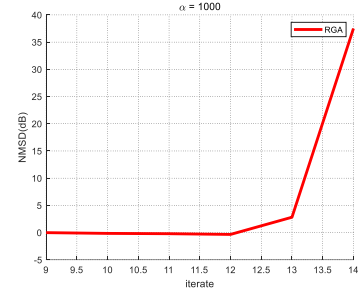

Fig. 2. NMSD simulation

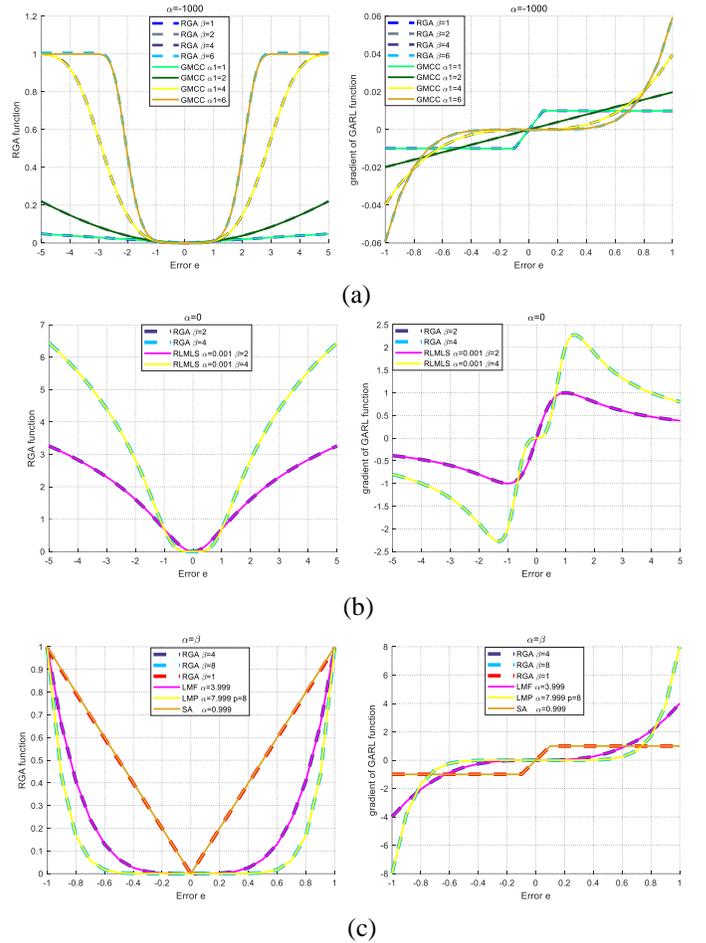

(a)

(b)

(c)

Fig. 3. Curves of different CFs and gradients

*B. Definition of NRGA function and kernel methods*

For the generalized Gaussian kernel to be derived naturally by the RGA function, from (7), we prefer the parameter $\alpha$ to be less than 0. Thus, the negative RGA (NRGA) function is expressed as

$$P_{NRGA}(e,b,\beta,\lambda) = \frac{b+\beta}{b} - k(e) \quad (9)$$

where $b \overset{def}{=} -\alpha > 0$, $k_{b,\beta}(e) = \frac{b+\beta}{b} \bigg/ \left(\frac{\lambda|e|^\beta}{b+\beta}+1\right)^{b/\beta}$.

Define two random variables $X$ and $Y$, and $x$, $y$ are drawn from joint pdf $f_{X,Y}(x,y)$. Substituting $e = x - y$ into (9) gives

$$\kappa_{\lambda,\beta}(x,y) = k_{b,\beta}(x-y) = \frac{b+\beta}{b} \bigg/ \left(\frac{\lambda|x-y|^\beta}{b+\beta}+1\right)^{b/\beta} \quad (10)$$

From the perspective of kernel functions, formula (10) satisfies Mercer's theorem when $\beta=2$. Consequently, the input data can be converted into a high-dimensional reproducing kernel Hilbert space $F_R$ through a nonlinear mapping $\phi_F$ induced by the kernel function

$$\kappa_{\lambda,2}(x,y) = \langle \phi_F(x), \phi_F(y)\rangle_{F_R} \quad (11)$$

here $\langle \cdot,\cdot \rangle_F$ is inner product in $F_R$.

Therefore, the NRGA kernel function, similar to the generalized Gaussian kernel, is a generalized form of the original kernel function with $\beta=2$. The specific properties of this function will be discussed in detail in the next section.

To describe its properties, we give two definitions of the NRGA entropy.

$$\rho(X,Y) = E\left[\kappa_{\lambda,\beta}(X-Y)\right] \quad (12)$$

$$\hat{\rho}(X,Y) = \frac{1}{N}\sum_{i=1}^{N}\kappa_{\lambda,\beta}(x_i - y_i) \quad (13)$$

here $X = \{(x_i)\}_{i=1}^{N}$, $Y = \{(y_i)\}_{i=1}^{N}$ are the finite amount of data, and $E(\cdot)$ denotes the expectation operator.

C. *The properties of NRGA entropy*

Some properties about NRGA entropy will be given in this subsection, there are several simple properties [40] will not be proved.

**Theorem 1.** Symmetric. $\rho(X,Y) = \rho(Y,X)$.

**Theorem 2.** The NRGA entropy $\rho(X,Y)$ is positive and bounded: $0 < \rho(X,Y) \leq \frac{b+\beta}{b}$. And it achieves to its maximum $\frac{b+\beta}{b}$, if and only if $X = Y$.

**Theorem 3.** The kernel function proposed in this paper does not have to satisfy the Mercer's condition. The kernel function $\kappa_{\lambda,\beta}(x,y)$ is positive define if and only if $0 < \beta \leq 2$.

**Proof.** First, we give the lemma of *positive definite* [see [41], page 434] and definition of *completely monotonic(c.m.)*.

*Lemma. 1:* Function $F(|x-x_i|)$ is a positive definite if and only if $F(\sqrt{|x-x_i|})$ is continuous and c.m.

*Definition. 1:* Let us call function $F(u)$ c.m. on $(0,\infty)$, provided that it is in $C^\infty(0,\infty)$ and satisfies the conditions

$$(-1)^k F^{(k)}(u) \geq 0, \ u \in (0,\infty), \ k=0,1,\ldots \quad (14)$$

To simplify the calculations, the NRGA kernel function is rewritten as

$$F(u) = m^b \quad (15)$$

where $m = \left(a/(\lambda u^\beta + a)\right)$.

From [see [42], page 3 (1.6)], if $f(x)$ is c.m., the function below is c.m.

$$f(\lambda x^\beta + a), \lambda \geq 0, a \geq 0 \text{ and } 0 < \beta \leq 1 \quad (16)$$

Clearly $\frac{1}{x}$ is c.m. and hence $m$ is c.m., when $0 < \beta \leq 1$. And, $F(u)$ is also c.m. by D.1 and Leibniz formula. Using Lemma.1, we can get the kernel $\kappa_{\lambda,\beta}(x,y)$ is positive define if and only if $0 < \beta \leq 2$.

**Theorem 4.** For $0 < \beta \leq 2$, $\rho(X,Y)$ is second-order statistic of the mapped feature space data.

**Proof.** When $0 < \beta \leq 2$, the kernel $\kappa_{\lambda,\beta}(x,y)$ is Mercer kernel, then we can have $\rho(X,Y) = E\left(\phi_F(X)^T \phi_F(Y)\right)$.

Introducing the CF $J_{NRGA\text{-cost}}$, between $X$ and $Y$ is defined as

$$J_{NRGA\text{-cost}}(X,Y) = \frac{b+\beta}{b} - \rho(X,Y) \quad (17)$$

the CF $J_{NRGA\text{-cost}} \geq 0$, and when $0 < \beta \leq 2$, $J_{NRGA\text{-cost}}$ can be denoted as

$$J_{NRGA\text{-cost}}(X,Y) = \frac{1}{2}E\left(\|\phi_F(X) - \phi_F(Y)\|^2\right) \quad (18)$$

which is a L2 loss in the feature space $F_R$ induced by the Mercer kernel $\kappa_{\lambda,\beta}(x,y)$.

**Theorem 5.** In the sample space, we give two vectors $X = (x_1, x_2, \ldots x_N)$ and $Y = (y_1, y_2, \ldots y_N)$. The function $IM_{NRGA}(X,Y) = (\kappa(0) - \rho(X,Y))^{1/2}$ defines a metric in the sample space.

**Proof.** The $IM_{NRGA}$ can be a metric if it meets the following properties:

1. Nonnegativity by Theorem 2.
2. Identity of indiscernibles. $IM_{NRGA}(X,Y) = 0$ if and only if $X = Y$ by Theorem 2.
3. Symmetric by Theorem 1.
4. Triangle inequality: $IM_{NRGA}(X,Z) \leq IM_{NRGA}(X,Y) + IM_{NRGA}(Y,Z)$. Based on $X$, $Y$, by nonlinear mapping $\phi_F$, we give two new vectors $\tilde{X} = [\phi_F(x_1); \phi_F(x_2); \ldots; \phi_F(x_N)]$ and $\tilde{Y} = [\phi_F(y_1); \phi_F(y_2); \ldots; \phi_F(y_N)]$ in Hilbert space $F_R^N$. The Euclidean distance $ED(\tilde{X}, \tilde{Y})$ can be represented as

$$ED(\tilde{X},\tilde{Y}) = \left(\langle(\tilde{X}-\tilde{Y}),(\tilde{X}-\tilde{Y})\rangle\right)^{1/2}$$
$$= \left(\langle\tilde{X},\tilde{X}\rangle - 2\langle\tilde{X},\tilde{Y}\rangle + \langle\tilde{Y},\tilde{Y}\rangle\right)^{1/2}$$
$$= \left(\sum_{i=1}^{N}\kappa_{\lambda,\beta}(x_i-x_i) + \sum_{i=1}^{N}\kappa_{\lambda,\beta}(y_i-y_i) + 2\sum_{i=1}^{N}\kappa_{\lambda,\beta}(x_i-y_i)\right)^{1/2}$$
$$= \left(2N\cdot(\kappa(0)-\rho(X,Y))\right)^{1/2}$$
$$= \sqrt{2N}\cdot IM_{NRGA}(X,Y) \qquad (19)$$

And,
$$IM_{NRGA}(X,Z) = ED(\tilde{X},\tilde{Z})/\sqrt{2N}$$
$$\leq ED(\tilde{X},\tilde{Y})/\sqrt{2N} + ED(\tilde{Y},\tilde{Z})/\sqrt{2N} \qquad (20)$$
$$= IM_{NRGA}(X,Y) + IM_{NRGA}(Y,Z)$$

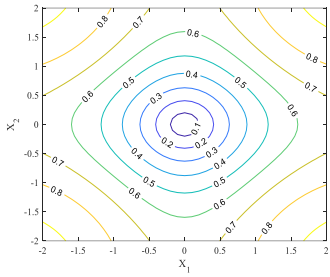

(a) $\beta=2$

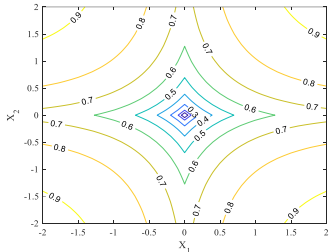

(b) $\beta=1$

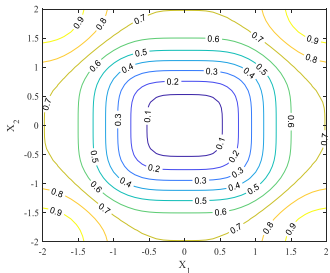

(c) $\beta=4$

Fig. 4. Contours of $IM_{NRGA}(X,0)$ in 2-D sample space

*Remark 3:* The contours of the induced metric $IM_{NRGA}$ under various parameters are depicted in Fig. 4. In Fig. 4(a), when the parameters are $b=1000$, $\beta=2$, and $\lambda=1$, the contour of $IM_{NRGA}$ at this time is similar to the correntropy induced metric (CIM). The induced metric $IM_{NRGA}$ behaves approximately as L2 norm for two points are close, and behaves as L1 norm for two points are further away. When the two points are particularly far away it behaves as L0 norm. As illustrated in Fig. 4(b) and (c), considering the parameters $\beta=1$ and $\beta=4$, the induced metric $IM_{NRGA}$ is similar to the generalized CIM (GCIM) at this moment. The evidence indicates that one variant of the metric space derived from the NRGA function closely resembles the GCIM, affirming the robust fitting performance of the NRGA function.

*D. Asymmetric problem*

The analysis in the previous subsections is only for symmetric error distributions. To further address the problem of asymmetric error distribution, based on the kernel function $\kappa_{\lambda,\beta}(x,y)$ defined in (9), the asymmetric kernel function is proposed as

$$\kappa_A(x,y) = \Delta_{\lambda_+\lambda_-}(e) = \begin{cases} \dfrac{b+\beta}{b}\Big/\left(\dfrac{\lambda_+|e|^{\beta}}{b+\beta}+1\right)^{b/\beta}, & e\geq 0 \\ \dfrac{b+\beta}{b}\Big/\left(\dfrac{\lambda_-|e|^{\beta}}{b+\beta}+1\right)^{b/\beta}, & e<0 \end{cases} \qquad (21)$$

where $\lambda_+$ and $\lambda_-$ correspond to the positive and negative parts of the error $e$, respectively. Symmetric function $\kappa_{\lambda,\beta}(x,y)$ and asymmetric function $\kappa_A(x,y)$ curves under different $\beta$ conditions are depicted in Fig. 5, and $b=100$. Obviously, asymmetric function $\kappa_A(x,y)$ will become the symmetric function $\kappa_{\lambda,\beta}(x,y)$ when $\lambda_+=\lambda_-$.

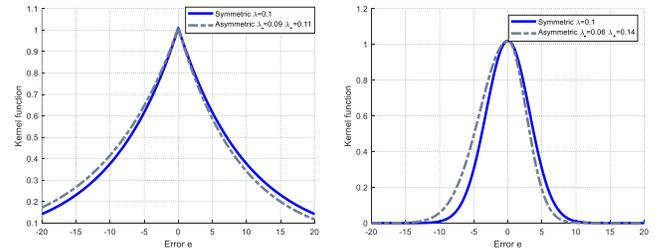

(a) $\beta=1$ and $\beta=2$

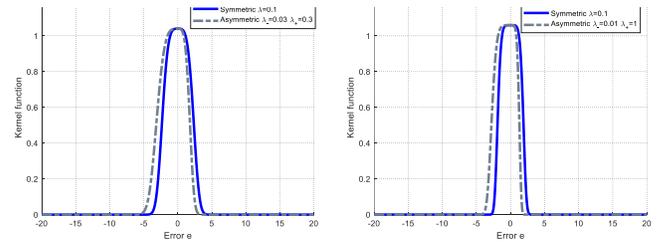

(b) $\beta=4$ and $\beta=6$

Fig. 5. Comparison of symmetric and asymmetric function

Additionally, it can be seen that the asymmetric NRGA entropy $\rho_A(X,Y)$ is positive and bounded in the range $0<\rho_A(X,Y)\leq \dfrac{b+\beta}{b}$, and it achieves its maximum if and only if $X=Y$.

## IV. THE PROPOSED ALGORITHMS

In this section, the derivation processes for the RGA algorithm and its variants, the NARGA and KRNRGA algorithm will be presented. The specific update processes are detailed in Table I, II, and III.

### A. The RGA algorithm

This flexibility in the shape of the CF provides better adaptation to changes in the noise environment. In the field of AF, the non-quadratic term of the error is generally applied to deal with generalized Gaussian noise, so more general robust generalized adaptive CF has been proposed as

$$J_{RGA-cost}(c) = E\left\{ \frac{|\alpha-\beta|}{\alpha}\left[\left(\frac{\lambda|e|^{\beta}}{|\alpha-\beta|}+1\right)^{\frac{\alpha}{\beta}}-1\right]\right\} \quad (22)$$

where $\alpha \in \mathbb{R}$, $\beta > 0$ are shape parameters and $\lambda > 0$ is a scale parameter.

Then, the gradient of $J_{RGA\text{-}cost}$ can be obtained as

$$\begin{aligned}\mathbf{g}_{RGA-cost}(c) &= \frac{\partial J_{RGA\text{-}cost}(c)}{\partial c} \\ &= E\left\{-\lambda|e|^{\beta-1}\text{sgn}(e)\left(\frac{\lambda|e|^{\beta}}{|\alpha-\beta|}+1\right)^{\frac{\alpha}{\beta}-1}x\right\}\end{aligned} \quad (23)$$

The instantaneous gradient $\hat{\mathbf{g}}_{RGA-cost}(w)$ of the RGA algorithm can be derived as

$$\hat{\mathbf{g}}_{RGA-cost}(c) = -\lambda|e|^{\beta-1}\text{sgn}(e)\left(\frac{\lambda|e|^{\beta}}{|\alpha-\beta|}+1\right)^{\frac{\alpha}{\beta}-1}x \quad (24)$$

Using the gradient descent method, the weight vector update formula is obtained as

$$\begin{aligned}c_{i+1} &= c_i - \mu\hat{\mathbf{g}}_{RGA-cost}(c) \\ &= c_i + \eta|e_i|^{\beta-1}\text{sgn}(e_i)\left(\frac{\lambda|e_i|^{\beta}}{|\alpha-\beta|}+1\right)^{\frac{\alpha}{\beta}-1}x_i\end{aligned} \quad (25)$$

where $\eta > 0$ is the step size parameter.

Overall, Table I details the exact steps of the RGA algorithm.

TABLE I
Summary of RGA Algorithm

| |
|---|
| Initialization: $c(0) = \mathbf{0}$ |
| Parameters: $\alpha$, $\beta$, $\lambda$, $\mu$ |
| For $i = 0, 1, 2$ |
| $\quad e_i = \tilde{d}_i - y_i$ |
| $\quad y_i = c_i^T x_i$ |
| $\quad \hat{\mathbf{g}}_{RGA-cost}(c) = -\lambda|e_i|^{\beta-1}\text{sgn}(e_i)\left(\frac{\lambda|e_i|^{\beta}}{|\alpha-\beta|}+1\right)^{\frac{\alpha}{\beta}-1}x_i$ |
| $\quad c_{i+1} = c_i - \mu\hat{\mathbf{g}}_{RGA-cost}(c)$ |
| END |

### B. The NARGA algorithm

To overcome the situation where conventional kernel functions may not adequately approximate PDF of the error distribution, the NARGA CF has been proposed as

$$J_{NARGA-cost}(e) = \frac{b+\beta}{b} - E\left(\Delta_{\lambda_+\lambda_-}(e)\right) \quad (26)$$

The instantaneous gradient $\hat{\mathbf{g}}_{NARGA-cost}(w)$ of the NARGA algorithm can be obtained as

$$\begin{aligned}\Lambda_{i+1} &= \Lambda_i - \mu\frac{\partial}{\partial \Lambda_i}\Delta_{\lambda_+\lambda_-}(e) \\ &= \Lambda_i - \mu\psi_{\lambda_+\lambda_-}(e)x_i\end{aligned} \quad (27)$$

where $\mu$ is step size parameter.

And the function $\psi_{\lambda_+\lambda_-}(e)$ can be given as

$$\psi_{\lambda_+\lambda_-}(e) = \begin{cases} -\lambda_+\text{sign}(e)|e|^{\beta-1}\left(\frac{\lambda_+|e|^{\beta}}{b+\beta}+1\right)^{\frac{-b-\beta}{\beta}}x_i, & e \geq 0 \\ -\lambda_-\text{sign}(e)|e|^{\beta-1}\left(\frac{\lambda_-|e|^{\beta}}{b+\beta}+1\right)^{\frac{-b-\beta}{\beta}}x_i, & e < 0 \end{cases} \quad (28)$$

It is evident that when $\lambda_+ = \lambda_-$, the NARGA CF degenerates into the NRGA CF. Additionally, the NARGA algorithm can be seen as a VSSLMS algorithm, denoted as

$$u_i = \begin{cases} -u\lambda_+|e|^{\beta-2}\left(\frac{\lambda_+|e|^{\beta}}{b+\beta}+1\right)^{\frac{-b-\beta}{\beta}}x_i, & e \geq 0 \\ -u\lambda_-|e|^{\beta-2}\left(\frac{\lambda_-|e|^{\beta}}{b+\beta}+1\right)^{\frac{-b-\beta}{\beta}}x_i, & e < 0 \end{cases} \quad (29)$$

To conclude, the specific steps of the NARGA algorithm are described in Table II.

TABLE II
Summary of NARGA Algorithm

| |
|---|
| Initialization: $c(0) = \mathbf{0}$ |
| Parameters: $b$, $\beta$, $\lambda_+$, $\lambda_-$, $\mu$ |
| For $i = 0, 1, 2$ |
| $\quad e_i = \tilde{d}_i - y_i$ |
| $\quad y_i = \Lambda_i^T x_i$ |
| $\quad \psi_{\lambda_+\lambda_-}(e) = \begin{cases} -\lambda_+\text{sign}(e)|e|^{\beta-1}\left(\frac{\lambda_+|e|^{\beta}}{b+\beta}+1\right)^{\frac{-b-\beta}{\beta}}x_i, & e \geq 0 \\ -\lambda_-\text{sign}(e)|e|^{\beta-1}\left(\frac{\lambda_-|e|^{\beta}}{b+\beta}+1\right)^{\frac{-b-\beta}{\beta}}x_i, & e < 0 \end{cases}$ |
| $\quad \Lambda_{i+1} = \Lambda_i - \mu\psi_{\lambda_+\lambda_-}(e)x_i$ |
| END |

## C. The Kernel Recursive NRGA Algorithm

In this subsection, the kernel recursive NRGA algorithm will be derived. First, using the NRGA function, the CF for kernel recursive least squares is represented as

$$J = \max_{c} \sum_{n=i}^{n+L-1} k_{b,\beta}(e_n) - \frac{1}{2}\gamma \|c_i\|^2 \quad (30)$$

where $e_n = d_n - c_i^T \phi_n$, and $\gamma$ is regularization factor. By taking the gradient of (30) with respect to $c_i$ and setting it to zero, the solution to the problem can be obtained as

$$c_i = \left[ \Phi_L \Omega_L \Phi_L^T + \frac{\gamma}{\lambda} I \right]^{-1} \Phi_L \Omega_L d_L \quad (31)$$

where

$$\Phi_L = [\phi_n, \phi_{n+1}, \ldots, \phi_{n+L-1}] = [\Phi_{L-1}, \phi_{n+L-1}] \quad (32)$$

$$d_L = [d_n, d_{n+1}, \ldots, d_{n+L-1}] = [d_{L-1}, d_{n+L-1}] \quad (33)$$

$$\Omega_L = diag\left[ |e_n|^{\beta-2} \left( \frac{\lambda |e_n|^\beta}{b+\beta} + 1 \right)^{-\frac{b+\beta}{\beta}}, |e_{n+1}|^{\beta-2} \left( \frac{\lambda |e_{n+1}|^\beta}{b+\beta} + 1 \right)^{-\frac{b+\beta}{\beta}}, \ldots, |e_{n+L-1}|^{\beta-2} \left( \frac{\lambda |e_{n+L-1}|^\beta}{b+\beta} + 1 \right)^{-\frac{b+\beta}{\beta}} \right] \quad (34)$$

Using matrix inversion lemma, (31) can be rewritten as

$$c_i = \Phi_L \left[ \Phi_L^T \Phi_L + \frac{\gamma}{\lambda} \Omega_L^{-1} \right]^{-1} d_L \quad (35)$$

Thus, according to (35), the weight vector $c_i$ can be clearly represented as

$$c_i = \Phi_L \Theta_L = \Phi_L Q_L d_L \quad (36)$$

And we have

$$\Theta_L = \left[ \Phi_L^T \Phi_L + \frac{\gamma}{\lambda} \Omega_L^{-1} \right]^{-1} d_L \quad (37)$$

$$Q_L = \left[ \Phi_L^T \Phi_L + \frac{\gamma}{\lambda} \Omega_L^{-1} \right]^{-1}$$

$$= \begin{bmatrix} \Phi_{L-1}^T \Phi_{L-1} + \frac{\gamma}{\lambda} \Omega_{L-1}^{-1} & h_L \\ h_L^T & \phi_{n+L-1}^T \phi_{n+L-1} + \frac{\gamma}{\lambda} \vartheta_L \end{bmatrix}^{-1} \quad (38)$$

where

$$h_L = \Phi_{L-1}^T \phi_{n+L-1}, \quad \vartheta_L = \left( |e_{n+L-1}|^{\beta-2} \left( \frac{\lambda |e_{n+L-1}|^\beta}{b+\beta} + 1 \right)^{-\frac{b+\beta}{\beta}} \right)^{-1} \quad (39)$$

Then, (38) can be further expressed as

$$Q_L^{-1} = \begin{bmatrix} Q_{L-1}^{-1} & h_L \\ h_L^T & \phi_{n+L-1}^T \phi_{n+L-1} + \frac{\gamma}{\lambda} \vartheta_L \end{bmatrix} \quad (40)$$

Using the block matrix inversion identity, we can have

$$Q_L = \varepsilon_L^{-1} \begin{bmatrix} \varepsilon_L Q_{L-1} + z_L z_L^T & -z_L \\ -z_L^T & 1 \end{bmatrix} \quad (41)$$

where $z_L = Q_{L-1} h_L$ and $\varepsilon_L = \phi_{n+L-1}^T \phi_{n+L-1} + \frac{\gamma}{\lambda} \vartheta_L - z_L^T h_L$.

Combining (37) and (41), we have

$$\Theta_L = Q_L d_L = \begin{bmatrix} \Theta_{L-1} - z_L \varepsilon_L^{-1} e_L \\ \varepsilon_L^{-1} e_L \end{bmatrix} \quad (42)$$

where $e_L = d_L - y_L = d_L - h_L^T \Theta_{L-1}$. Moreover, as the training data evolves, the complexity escalates, necessitating the use of the ALD criterion [34] to control the growth of the kernel matrix. Overall, Table III details the exact steps of the KRNRGA algorithm.

TABLE III
Summary of KRNRGA Algorithm

---

Initialization: $Q_1 = \left[ \frac{\gamma}{\lambda} + \kappa_{\lambda,\beta}(x_1, x_1) \right]^{-1}$, $\Theta_1 = Q_1 d_1$

Parameters: $b$, $\beta$, $\lambda$, $\gamma$

While $\{d_n, x_n\} \neq \emptyset$ do

  Computation:

  $h_L = \Phi_{L-1}^T \phi_{n+L-1} = \left[ \kappa_{\lambda,\beta}(x_{n+L-1}, x_n), \ldots, \kappa_{\lambda,\beta}(x_{n+L-1}, x_{n+L-2}) \right]^T$

  $y_L = h_L^T \Theta_{L-1}$

  $e_L = d_L - y_L$

  Compute $z_L$, $\vartheta_L$, $\varepsilon_L$

  Update $Q_L$ and $\Theta_L$

END

---

## V. STABILITY AND STEADY-STATE PERFORMANCE

### A. Mean convergence

First, we provided some plausible assumptions to simplify the subsequent analysis [43], [44], [45].

**Assumption 1:** The weight error $\Delta c_i \stackrel{def}{=} c_o - c_i$, is uncorrelated with $d_i$ and $x_i$.

**Assumption 2:** The input vector $x_i$ and desired signal $d_i$ are zero-mean joint Gaussian distributed.

Considering A1, (9) can be rewritten as

$$\Delta c_{i+1} = \Delta c_i - \eta |e_i|^{\beta-1} \text{sgn}(e_i) \left( \frac{\lambda |e_i|^\beta}{|\alpha - \beta|} + 1 \right)^{\frac{\alpha}{\beta}-1} x_i \quad (43)$$

Then, taking the mathematical expectations on (43), we get

$$E[\Delta c_{i+1}] = E[\Delta c_i] - \eta E[f(e_i) x_i] \quad (44)$$

where

$$f(e_i) = |e_i|^{\beta-1} \text{sgn}(e_i) \left( \frac{\lambda |e_i|^\beta}{|\alpha - \beta|} + 1 \right)^{\frac{\alpha}{\beta}-1} \quad (45)$$

Using the Lemma 1 of [45] and A2, $E[f(e_i) x_i]$ can be represented as

$$E[f(e_i)\mathbf{x}_i] = \phi_G(e_i) E[e_i \mathbf{x}_i] \quad (46)$$

where

$$\phi_G(e_i) = \frac{E[f(e_i)e_i]}{E[e_i^2]} \quad (47)$$

Substituting (46) into (44) yields

$$E[\Delta \mathbf{c}_{i+1}] = E[\Delta \mathbf{c}_i] - \eta \phi_G(e_i) E[\mathbf{x}_i \mathbf{x}_i^T] E[\Delta \mathbf{c}_i]$$
$$= (\mathbf{I} - \eta \phi_G(e_i) \mathbf{R}_x) E[\Delta \mathbf{c}_i] \quad (48)$$

where $\mathbf{R}_x = E[\mathbf{x}_i \mathbf{x}_i^T]$ is autocorrelation matrix of input vector.

The step size $\mu$ that ensures stable operation of the RGA algorithm gives by (48)

$$0 < \mu < \frac{2}{\phi_G(e_i) tr(\mathbf{R}_x)} \quad (49)$$

here $tr(\cdot)$ is the trace of the matrix.

*B. Steady-State MSD*

The steady-state MSD of the RGA algorithm is calculated in this subsection. For simplicity of calculation, some common assumptions are presented as below [45][46]

**Assumption 3**: $\|\mathbf{x}_i\|^2$ and $f^2(e_i)$ are asymptotically uncorrelated.

**Assumption 4**: Both the input vector $\mathbf{x}_i$ and output noise $v_i$ are Gaussian processes with zero mean, independent and identically distributed. $\mathbf{x}_i$ and $v_i$ are independent of each other.

**Assumption 5**: The *apriori* error $e_a$ is zero mean and is independent of the output noise $v_i$.

The *apriori* error, and systematic error are defined as

$$e_{a,i} = \Delta \mathbf{c}_i^T \mathbf{x}_i \quad (50)$$
$$e_i = e_{a,i} + v_i \quad (51)$$

Then, we can obtain the corresponding variance as

$$\sigma_e^2 = E(e_{a,i}^2) + \sigma_v^2 \quad (52)$$

where $\sigma_v^2$ is the variance of output noise.

From the energy relation, and combing (43)(45) gives

$$E[\|\Delta \mathbf{c}_{i+1}\|^2] = E[\|\Delta \mathbf{c}_i\|^2] - 2\eta E[f(e_i) e_{a,i}]$$
$$+ \eta^2 E[f^2(e_i) \|\mathbf{x}_i\|^2] \quad (53)$$

Under A3~A5, (53) can be further derived as

$$E[\|\Delta \mathbf{c}_{i+1}\|^2] = E[\|\Delta \mathbf{c}_i\|^2] - 2\eta \phi_G(e_i) E[\|\Delta \mathbf{c}_i\|^2 \mathbf{R}_x]$$
$$+ \eta^2 \phi_U(e_i) E[\|\mathbf{x}_i\|^2] \quad (54)$$

where $\phi_G(e_i) = E(e_i f(e_i))/E(e_i^2)$ and $\phi_U(e_i) = E(f^2(e_i))$.

Moreover, at the steady state gives

$$\lim_{i \to \infty} E(\|\Delta \mathbf{c}_{i+1}\|^2) = \lim_{i \to \infty} E(\|\Delta \mathbf{c}_i\|^2) \quad (55)$$

Substituting (55) into (54) at the steady state yields

$$\lim_{i \to \infty} \phi_G(e_i) \lim_{i \to \infty} E[\|\Delta \mathbf{c}_i\|^2 \mathbf{R}_x] = \frac{\eta}{2} E[\|\mathbf{x}_i\|^2] \lim_{i \to \infty} \phi_U(e_i) \quad (56)$$

The s excess mean-square error (EMSE) at steady state can be represented as

$$\psi = \lim_{i \to \infty} E(e_{a,i}^2) = \lim_{i \to \infty} E(|\Delta \mathbf{c}_i^T \mathbf{x}_i|^2) = \lim_{i \to \infty} E(\|\Delta \mathbf{c}_i\|^2 \mathbf{R}_x) \quad (57)$$

Substituting (57) into (56) provides

$$\psi = \frac{\eta}{2} E[\|\mathbf{x}_i\|^2] \frac{\lim_{i \to \infty} \phi_U(e_i)}{\lim_{i \to \infty} \phi_G(e_i)} = \frac{\eta}{2} tr(\mathbf{R}_x) \frac{\lim_{i \to \infty} \phi_U(e_i)}{\lim_{i \to \infty} \phi_G(e_i)} \quad (58)$$

In the situation where the output noise is Gaussian noise $(\beta = 2)$ and the $e_i$ is very small at steady state, we have

$$\lim_{i \to \infty} \phi_G(e_i) = \lim_{i \to \infty} \frac{E[f(e_i)e_i]}{E[e_i^2]}$$
$$= \lim_{i \to \infty} \frac{E\left(e_i \left(\frac{\lambda e_i^2}{|\alpha - 2|} + 1\right)^{\frac{\alpha}{2}-1} e_i\right)}{E[e_i^2]} \quad (59)$$
$$= 1$$

$$\lim_{i \to \infty} \phi_U(e_i) = \sigma_e^2 \quad (60)$$

Substituting (52), (59), and (60) into (58) gives

$$\psi = \frac{\eta}{2} tr(\mathbf{R}_x) \sigma_e^2 = \frac{\eta tr(\mathbf{R}_x) \sigma_v^2}{2 - \eta tr(\mathbf{R}_x)} \quad (61)$$

Then, we have

$$MSD(\infty) = \frac{L}{tr(\mathbf{R}_x)} \psi = \frac{L \eta \sigma_v^2}{2 - \eta tr(\mathbf{R}_x)} \quad (62)$$

VI. COMPUTATIONAL COMPLEXITY

The computational complexity of different algorithms is analyzed in this subsection and summarized in Table IV.

From Table IV, the RGA algorithm has increased its computational complexity compared to the GMCC algorithm. But this increase depends on the selection of parameters $\alpha$ and $\beta$. If smaller values of $\alpha$ and $\beta$ are chosen, the complexity of the RGA algorithm does not have a significant increase compared to GMCC.

TABLE IV
Computational Complexity

| Algorithm | $+/-$ | $\times/\div$ | Nonlinear |
|---|---|---|---|
| LMS | 2L | 2L+1 | 0 |
| MCC | 2L | 2L+6 | 1 |
| LMLS | 2L+1 | 2L+3 | 1 |
| GMCC | 2L | $2L+2+2\alpha$ | 4 |
| SA | 2L | 2L | 1 |
| RLMLS | 2L+1 | 2L+9 | 2 |
| RGA | 2L+4 | $2L+1+2\beta+\alpha/\beta$ | 4 |

TABLE V
Simulation Parameters of Symmetric Noise Distribution

| Algorithms | Parameters | | | | | |
|---|---|---|---|---|---|---|
| | Noise(1) | Noise(2) | Noise(3) | Noise(4) | Noise(5) | Noise(6) |
| LMS | $\mu=0.0027$ | $\mu=0.0027$ | $\mu=0.0027$ | $\mu=0.0027$ | $\mu=0.0027$ | $\mu=0.0027$ |
| MCC | $\mu=0.03\ \sigma=1$ | $\mu=0.03\ \sigma=1$ | $\mu=0.03\ \sigma=1$ | $\mu=0.03\ \sigma=1$ | $\mu=0.03\ \sigma=1$ | $\mu=0.009\ \sigma=1$ |
| LMLS | $\mu=0.02\ \gamma=1$ | $\mu=0.02\ \gamma=1$ | $\mu=0.02\ \gamma=1$ | $\mu=0.02\ \gamma=1$ | $\mu=0.02\ \gamma=1$ | $\mu=0.009\ \gamma=1$ |
| GMCC | $\mu=0.0055\ \lambda=0.01\ \alpha=2$ | $\mu=0.0055\ \lambda=0.01\ \alpha=2$ | $\mu=0.007\ \lambda=0.01\ \alpha=1$ | $\mu=0.0012\ \lambda=0.01\ \alpha=4$ | $\mu=0.0012\ \lambda=0.01\ \alpha=6$ | $\mu=0.01\ \lambda=0.1\ \alpha=3$ |
| RGA | $\mu=0.45\ \lambda=0.01\ \alpha=-100\ \beta=2.1$ | $\mu=0.45\ \lambda=0.01\ \alpha=-100\ \beta=2.1$ | $\mu=0.57\ \lambda=0.01\ \alpha=-100\ \beta=1.5$ | $\mu=0.018\ \lambda=0.01\ \alpha=-1000\ \beta=6$ | $\mu=0.033\ \lambda=0.01\ \alpha=-1000\ \beta=8$ | $\mu=0.065\ \lambda=0.1\ \alpha=-100\ \beta=2.2$ |
| LMF | $\mu=0.00035$ | | | | | |
| SA | $\mu=0.009$ | $\mu=0.009$ | $\mu=0.009$ | $\mu=0.009$ | $\mu=0.009$ | $\mu=0.005$ |
| RLMLS | $\mu=0.05\ \gamma=1$ | $\mu=0.05\ \gamma=1$ | $\mu=0.05\ \gamma=1$ | $\mu=0.05\ \gamma=1$ | $\mu=0.05\ \gamma=1$ | $\mu=0.05\ \gamma=1$ |

TABLE VI
Simulation Parameters of Asymmetric Noise Distribution

| Algorithms | Parameters | | |
|---|---|---|---|
| | Noise(7) | Noise(8) | Noise(9) |
| MCC | $\mu=0.018\ \sigma=1$ | $\mu=0.018\ \sigma=1$ | $\mu=0.01\ \sigma=1$ |
| MACC | $\mu=0.018\ \sigma_+=1.3\ \sigma_-=1.6$ | $\mu=0.018\ \sigma_+=2.6\ \sigma_-=0.8$ | $\mu=0.025\ \sigma_+=3\ \sigma_-=1.2$ |
| GMCC | $\mu=0.018\ \beta=1\ \alpha=2$ | $\mu=0.018\ \beta=1\ \alpha=2$ | $\mu=0.01\ \beta=1\ \alpha=2$ |
| RGA | $\mu=0.5\ \lambda=0.01\ \alpha=-100\ \beta=1.5$ | $\mu=0.35\ \lambda=0.01\ \alpha=-100\ \beta=1.5$ | $\mu=0.5\ \lambda=0.01\ \alpha=-100\ \beta=1.56$ |
| GMACC | $\mu=0.018\ \sigma_+=1.3\ \sigma_-=1.6\ \alpha=2$ | $\mu=0.018\ \sigma_+=2.6\ \sigma_-=0.8\ \alpha=2$ | $\mu=0.025\ \sigma_+=3\ \sigma_-=1.2\ \alpha=2$ |
| NARGA | $\mu=0.08\ \lambda_+=0.061\ \lambda_-=0.6\ \alpha=-100\ \beta=1.5$ | $\mu=0.105\ \lambda_+=0.036\ \lambda_-=0.35\ \alpha=-100\ \beta=1.5$ | $\mu=0.34\ \lambda_+=0.5\ \lambda_-=0.06\ \alpha=-100\ \beta=1.56$ |

TABLE VII
Simulation Parameters of Time Series Prediction

| Algorithms | Parameters | |
|---|---|---|
| | Noise(10) | Noise(11) |
| KRNRGA | $\gamma=10^{-1}\ \alpha=-200\ \beta=2.2\ \lambda=1.1$ | $\gamma=10^{-1}\ \alpha=-10\ \beta=2.1\ \lambda=1$ |
| KRGMCC | $\gamma=10^{-1}\ \alpha=2\ \sigma=2.7$ | $\gamma=10^{-1}\ \alpha=2\ \sigma=2.3$ |
| KRLS | $\gamma=10^{-1}$ | $\gamma=10^{-1}$ |

When $\alpha=2$, the computational complexity of the GMCC algorithm is comparable to the MCC algorithm. Moreover, although the computational complexity of the RGA algorithm is raised, utilizing the RGA algorithm ensures that the cost function does not have to be replaced when dealing with different noise environments, but only a simple adjustment of the algorithm parameters is required. In addition, according to the simulation part, the RGA algorithm obtains optimal performance in different noise environments. The computational complexity of the NARGA algorithm is same as the RGA algorithm because its computational process does not change, except that its kernel parameters take different values at different error intervals. Finally, since the KRNRGA algorithm has a significantly higher computational complexity than the AF algorithm, we do not discuss it here either.

## VII. SIMULATION

The superior performance of the RGA, NARGA, and KRNRGA algorithms proposed in this paper will be thoroughly validated. First, the following types of noise distributions will be considered:
(1) Zero-mean Gaussian noise with variance 1.
(2) Zero-mean Gaussian noise with variance 1 combined with impulsive noise generated by a Bernoulli process with variance 1000.
(3) Zero-mean Laplace noise with variance 2 combined impulsive noise with variance 1000.
(4) Binary noise over $[-2,2]$ with probability $\Pr\{x=-2\}=\Pr\{x=2\}=0.5$ combined impulsive noise with variance 1000.
(5) Uniform noise over $[-\sqrt{2},\sqrt{2}]$ combined impulsive noise with variance 1000.
(6) Generalized Gaussian Distribution (GGD) noise [22] with $\alpha=3$, $\beta=0.3$ and combined impulsive noise with variance 1000.
(7) Mixed Gaussian noise with $v_i=(1-m_i)H_i+m_iZ_i$ where $m_i$ is a binary process with $\Pr\{m_i=1\}=p$, $\Pr\{m_i=0\}=1-p$, and $0\le p\le 1$ is the probability of occurrence. $H_i$ and $Z_i$ are zero-mean Gaussian distributions with variances 1 and 400, respectively.
(8) Mixed Gaussian noise $v_i=(1-m_i)M_i+m_iZ_i$, where $M_i=A_i+B_i$. $A_i$ and $B_i$ are zero-mean Gaussian distributions with variances 0.8 and 8, respectively.
(9) $F$-distribution noise with $F(5,14)$.
(10) Zero-mean Gaussian noise with variance 0.1.
(11) Zero-mean Rayleigh noise with $v(t)=\exp(-t^2/2\sigma^2)(t/\sigma^2)$, where $\sigma=1.5$.

### A. Parameter selection

The effect of the parameters $\alpha$ and $\beta$ on the algorithm was analyzed in detail in Section III, so in this subsection we will analyze the effect of the parameter $\lambda$ on the algorithm.

First, the noise distribution (1) is selected as the background noise. All parameters except $\lambda$ follow the RGA algorithm parameter settings under the noise distribution (1) listed in Table V. Fig.6 shows the effect of different $\lambda$ on the RGA algorithm under this condition.

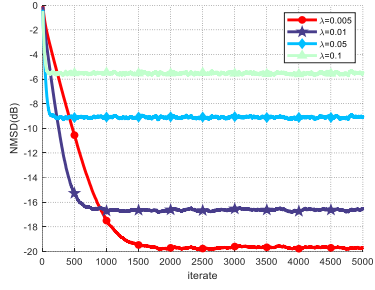

Fig. 6. Comparison of simulation results under different $\lambda$

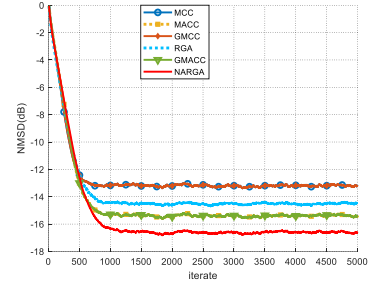

(a) Noise distribution (7)

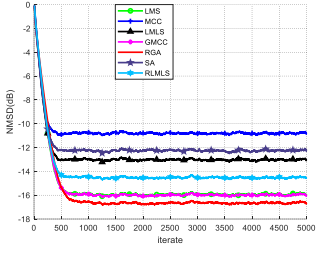
(a) Noise distribution (1)

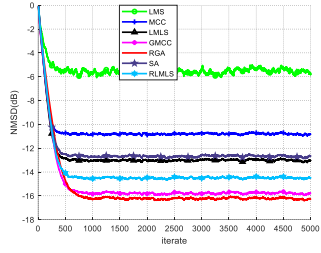
(b) Noise distribution (2)

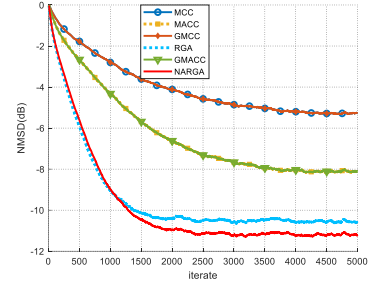

(b) Noise distribution (8)

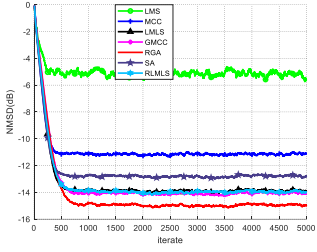
(c) Noise distribution (3)

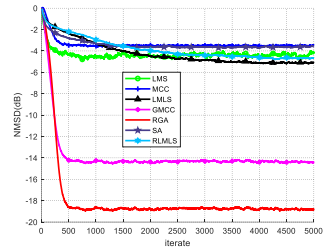
(d) Noise distribution (4)

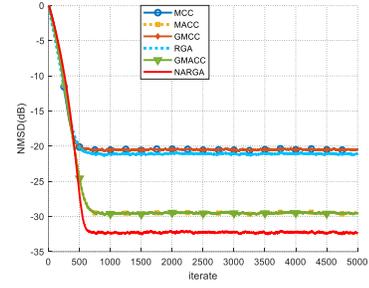

(c) Noise distribution (9)

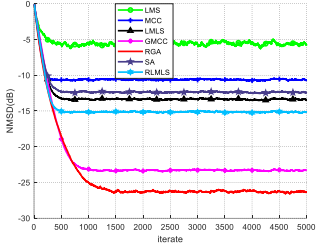
(e) Noise distribution (5)

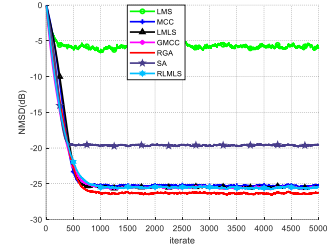
(f) Noise distribution (6)

Fig. 7. Performance comparison for different algorithms

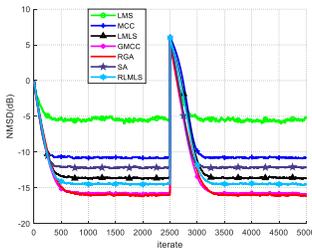
(a) Noise distribution (2)

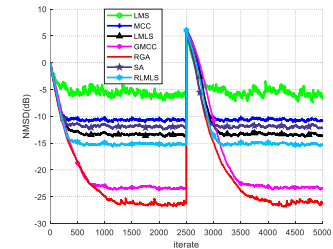
(b) Noise distribution (5)

Fig. 8. Tracking Performance comparison

Fig. 9. Performance comparison for different algorithms under asymmetric noise environment

From Fig. 6, it can be seen that when $\lambda$ becomes larger, the convergence speed of the algorithm is accelerated, but the steady-state error is larger; on the contrary, the convergence speed is slower and the steady-state error is smaller. In addition, when the value of $\lambda$ is too large, the convergence speed is not significantly improved, but the steady-state MSD increases significantly.

### B. System Identification

The performance of the RGA algorithm is demonstrated by averaging the simulated values over 1000 separate Monte Carlo (MC) runs under different background noise in this subsection. Unless otherwise indicated, the filter order is $L = 9$, and the sampling points is $N = 5000$. The parameters for the different algorithms are chosen to ensure that they have the same initial convergence rate to guarantee a fair comparison. The performance indicator normalized MSD (NMSD) of the algorithm is given as

$$NMSD = 10\log\left[ \left\| \mathbf{c}_i - \mathbf{c}_o \right\|^2 / \left\| \mathbf{c}_o \right\|^2 \right] \quad (63)$$

*a) Symmetric Noise Distribution*

The superiority of the RGA algorithm's performance in symmetric noise environments will be demonstrated in this subsection. Specifically, the noise conditions are (1) through (6) as mentioned in the previous section, with the corresponding simulation figures being Fig .7(a) through (f).

In Fig. 7(a) through (f), all algorithms fitted in Part A of Section III are compared under different noise environments. In noise distributions (2)-(6), due to the influence of impulsive noise, the LMF diverges and is therefore not included in the comparison. By adjusting the RGA parameters, the robustness of the algorithm can be adapted to different noise distributions. This adjustment ensures the superior performance of the RGA algorithm. As depicted in Fig. 7(a) to (c), and Fig. 7(f), the RGA algorithm only has a slight improvement compared to the other competing algorithms in noise environments (1) to (3), and (6). However, as depicted in Fig. 7(d) to (e), the performance of the RGA algorithm is significantly superior to the other competing algorithms in noise environments (4) through (5). This performance advantage arises from the scenario that existing algorithms cannot achieve optimal performance merely by parameter adjustments in varying noise environments. In contrast, the RGA function secures optimal performance through its exceptional fitting capability. The parameters used in this simulation are shown in Table V.

*b) Tracking performance*

In this subsection, the tracking performance of the RGA algorithm is verified under different noise distributions. In Fig.8, the unknown system weight vector $c_o$ mutates to $-c_o$ in 2500th input samples. We will select noise distribution (2) and noise distribution (5) as the background noise, respectively. The parameter selection is same as in Table V, corresponding to the noise environment.

As shown in Fig. 8(a) and (b), which correspond to noise distributions (2) and (5), under Gaussian noise, the RGA algorithm provides only a slight performance advantage; however, under uniform noise, the tracking performance of the RGA algorithm significantly outperforms other competing algorithms.

*c) Asymmetric Noise Distribution*

The superiority of the NARGA algorithm's performance in asymmetric noise environments will be validated in this subsection. In particular, the noise conditions are given by (7) through (9) as mentioned in the preceding section, with the corresponding simulation diagrams labeled as Fig. 9(a) through (c).

Since $a$ is set to 2, the simulation curves for the GMCC and MCC algorithm overlap in Fig. 9(a) through (c), which correspond to noise distributions (7) through (9), while the simulation curves for the GMACC and MACC algorithm also overlap. As shown in Fig. 9(a) through (c), the Steady-State MSD of the NARGA algorithm is significantly lower than other competing algorithms, thereby validating its effectiveness. Additionally, in the noise environment (8), the performance of the RGA algorithm is also superior to the MACC and GMACC algorithm, demonstrating the algorithm's strong adaptability to asymmetric noise distributions. The purpose of the MACC algorithm is to better fit the PDF of asymmetric error distribution, addressing the issue where the symmetric Gaussian kernel fails to fit asymmetric error PDF well, potentially leading to severe performance degradation [30]. For the NARGA algorithm, it has a better capability to fit asymmetric error distributions compared to traditional maximum correntropy, providing the algorithm with superior performance. The parameters used in this simulation are shown in Table VI.

### C. Time Series Prediction

This subsection employs a real Chua's circuit system to generate chaotic time series, providing a basis for assessing the nonlinear learning capability of the KRNRGA algorithm. The system and schematic diagram of Chua's circuit are illustrated in Fig. 10 and Fig. 11. The system of Chua's circuit is designed as

$$\begin{cases} \dot{u}_1 = \dfrac{u_2}{RC_1} - \dfrac{u_1}{RC_1} - \dfrac{\varphi(u_1)}{C_1} \\ \dot{u}_2 = \dfrac{u_1}{RC_2} - \dfrac{u_2}{RC_2} - \dfrac{I_{\tilde{L}}}{C_1} \\ I_{\tilde{L}} = -\dfrac{u_2}{\tilde{L}} \end{cases} \quad (64)$$

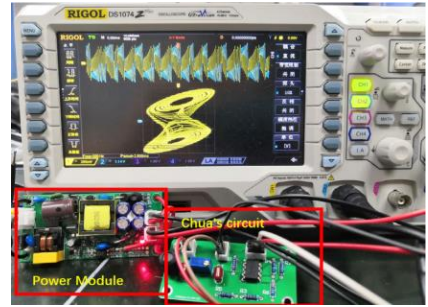

Fig .10. Chua's circuit system

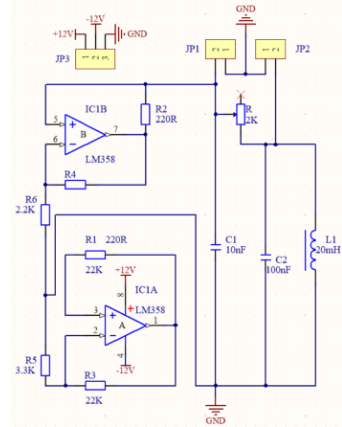

Fig. 11. Chua's circuit system schematic diagram

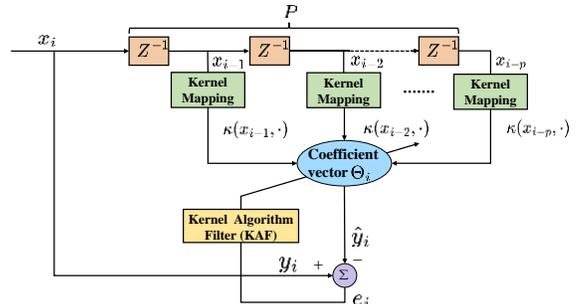

Fig. 12. Schematic diagram of the KAF algorithm applied to time series

where $I_{\tilde{L}}$, $u_1$ and $u_2$ denote the current of inductor $\tilde{L}$, and the voltage of capacitors $C_1$, $C_2$. $\varphi(u_1)$ is a segmented continuous function depicting the diode voltage-ampere characteristic. The voltage across capacitor $C_1$ is sampled for time series prediction, with the latest voltage value predicted based on the preceding five voltage values. The first 3000 noise contaminated samples are used to build the training set, while 100 additional samples constitute the test set to evaluate the performance of various algorithms. The KAF algorithm application with time series prediction can be represented as in Fig. 12. W+here $Z^{-1}$ is delay unit, $x_{i-1}$, $x_{i-2}$,…, $x_{i-p}$ are the input data, $\kappa(x_{i-1},\cdot)$, $\kappa(x_{i-2},\cdot)$,…, $\kappa(x_{i-p},\cdot)$ are the input data after kernel mapping, $y_i$ is the desired output data and is identical to the current data $x_i$. $\hat{y}_i$ is the estimation, which is obtained by multiplying $\kappa(x_{i-1},\cdot)$, $\kappa(x_{i-2},\cdot)$,…, $\kappa(x_{i-p},\cdot)$ with the coefficient vector $\Theta_i$. The difference between $y_i$ and $\hat{y}_i$ defined as error $e_i$. By continuously reducing the $e_i$, the KAF can continuously update the $\Theta_i$. Finally, we can implement time series prediction by utilizing first $P$ data. First kernel mapping is performed on these $P$ data, and then multiplying by the trained coefficient vector $\Theta_i$ to predict the $P+1 th$ data. The performance indicator mean square error (MSE) of the algorithm is given as

$$MSE = \frac{1}{N}\sum_{i=1}^{N} e_i^2 \quad (65)$$

As shown in Fig. 13(a) and (b), which correspond to noise distributions (10) and (11), the KRNRGA algorithm exhibits better performance over the other algorithms. This demonstrates that in real-world chaotic sequence prediction scenarios, the KRNRGA algorithm achieves optimal performance across different noise conditions, owing to its exceptional fitting capability. The parameters used in this simulation are shown in Table VII.

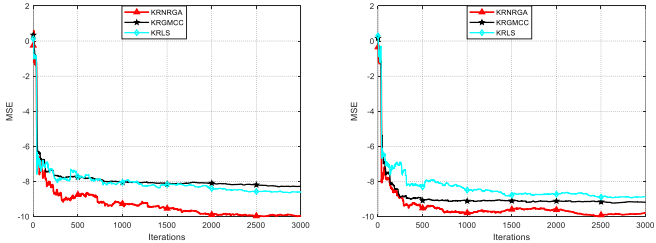

(a) Noise distribution (10)  (b) Noise distribution (11)

Fig. 13. Performance comparison for time series prediction

### D. Theoretical Verification

In this subsection, the theoretical steady-state MSD values computed for the RGA algorithm will be validated through computer simulations.

In Fig. 14, the noise environment is Gaussian noise with different variances. As indicated in Fig. 14 the theoretical steady-state MSD computed for the RGA algorithm fits very well with the values, which proves the correctness of the theoretical values calculations.

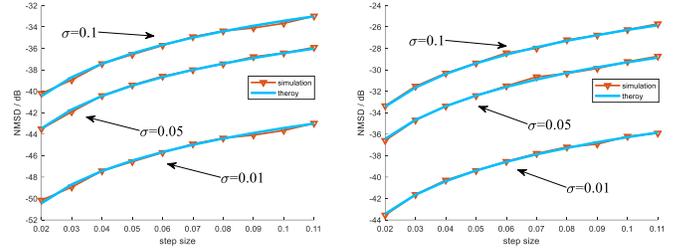

(a) $\alpha=-1000$ $\beta=2$ $\lambda=0.01$  (b) $\alpha=-1000$ $\beta=2$ $\lambda=0.05$

Fig. 14. Theoretical steady-state MSD verification

## VIII. CONCLUSION

In conclusion, to address the issue where different algorithms often fail to maintain their performance advantages when the noise environment changes, the RGA algorithm is proposed. Additionally, the NARGA algorithm is developed to address asymmetric error distribution problems, and the KRNRGA algorithm is designed for the time-series prediction problem of Chua's circuit The proposed algorithms exhibit strong adaptability to various noise environments, achieving performance advantages under different background noises by simply adjusting parameters. Furthermore, the NRGA algorithm can be considered an extension of the NRGA kernel, which has properties similar to the generalized Gaussian kernel. The properties of the NRGA entropy are analyzed in detail in this paper. Lastly, the convergence of the RGA algorithm and its steady-state MSD are thoroughly analyzed. Detailed simulations demonstrate that the proposed algorithms generally outperform other competing algorithms in various noise environments.